\documentclass[10pt, conference, a4paper]{IEEEtran}
\usepackage{xspace,amsmath,amssymb,amsfonts,epsfig,syntonly,amsthm}
\usepackage{cite,bm,color,url,textcomp}
\usepackage{algorithmicx,algorithm,algpseudocode}
\usepackage{epstopdf}
\usepackage{empheq}
\usepackage{geometry}
\long\def\symbolfootnote[#1]#2{\begingroup
\def\thefootnote{\fnsymbol{footnote}}
\footnote[#1]{#2}\endgroup}

\renewcommand{\thefootnote}{}

\begin{document}
\title{Two-Way Energy Trading and Online Planning for Fifth-Generation Communications with Renewables}
%\vspace{-0.8cm}
\author{\IEEEauthorblockN{Xiaojing Chen$^{1,3}$, Xin Wang$^1$, Wei Ni$^2$, and Iain B. Collings$^3$}
\IEEEauthorblockA{$^1$Key Lab of EMW Information (MoE), Dept. of Commun. Sci. \& Engr., Fudan University, Shanghai, China
      }
\IEEEauthorblockA{$^2$CSIRO, Sydney, NSW 2122, Australia
}
\IEEEauthorblockA{$^3$School of Engineering, Macquarie University, Sydney, NSW 2122, Australia
}
}

\maketitle
\footnotetext{Work in this paper was supported by the National Natural Science Foundation of China grants 61671154 and the Innovation Program of Shanghai Municipal Education Commission. }

\begin{abstract}
Future fifth-generation (5G) cellular networks, equipped with energy harvesting devices, are uniquely positioned to closely interoperate with smart grid. New interoperable functionalities are discussed in stochastic two-way energy trading and online planning to improve efficiency and productivity. Challenges lie in the unavailability of a-priori knowledge on future wireless channels, energy pricing and harvesting. Lyapunov optimization techniques are utilized to address the challenges and stochastically optimize energy trading and planning. Particularly, it is able to decouple the optimization of energy trading and planning during individual time slots, hence eliminating the need for joint optimization across a large number of slots.

\end{abstract}

\section{Introduction}\label{intro}

Communication network and smart grid have been independently evolving despite both aiming to substantially improve productivity, efficiency and sustainability.
On one hand, the upcoming fifth-generation (5G) cellular networks are anticipated to be densely deployed with a significantly reduced coverage area per cell. This is due to explosively increasing mobile traffic and propagation-unfriendly high-frequency spectrum which only avails for communication purpose \cite{Agiwal2016Next}.
Unfortunately, the overall energy consumption of base stations (BSs) would become increasingly overwhelming, and contribute adversely to the reduction of global carbon dioxide emissions.
On the other hand, smart grid is expected to usher in controllability, interoperability and sustainable exploitation of renewable energy sources, or renewables (RES) for short. Equipped with smart meters, smart grid is able to embrace brand new functionalities such as distributed energy generation, two-way energy flows, energy trading and redistribution, and energy demand management \cite{Fang2012Smart}.

This paper explores the potential of co-evolution of communication system and smart grid, as is becoming prominently important due to the fact that BSs are increasingly equipped with energy harvesting capabilities, such as solar panels and wind turbines~\cite{Li2015Decentralized}, for economical and ecological purposes. RES (e.g., up to 10,000 KW per BS) can be harvested to supplement persistent supplies from the conventional power grid and power cellular systems \cite{Han2014Powering}. Manufacturers, such as Ericsson, Huawei, and Vodafone, have started developing ``green'' BSs that can be powered jointly by persistent supplies and harvested RES \cite{Li2015Decentralized}.
%Cellular networks are also uniquely positioned to co-evolve with smart grid. The sheer scale and ubiquity of cellular networks can result in a significant amount of energy, either purchased off the grid or harvested from ambient environments. The close control of cellular networks alsp resembles to that of the smart grid, providing efficient energy redistribution and effective price negotiation with the smart grid \cite{Xu15}.

Under the co-evolution of communication system and smart grid, traditional energy consumers, such as BSs, can potentially become an integral part of the grid, generate and redistribute energy, and trade energy with the grid through the smart meters.
The BSs can purchase energy off the grid while in shortage of RES, and sell extra energy back to the grid when RES are in abundance \cite{Fang2012Smart}. Such two-way energy trading (TWET) allows extra energy to be redistributed for environmental benefits and financial gains of 5G.
%This helps balance energy load and relieve pressure on the grid, and hence improve the reliability and sustainability of the grid.
%
%\textit{Dynamic energy pricing:} As a result of intermittent renewable energy and TWET, energy prices are expected to exhibit strong dynamics in smart grid. The dynamic pricing is important to regulate the energy demands, and encourage users such as 5G networks to consume energy wisely and efficiently. The prices of both selling and buying energy, fluctuate along the time to reflect the real-time energy demand and supply availability.
%
Moreover, the energy prices of the smart grid can diverge contractually over different timescales, i.e., real-time price (e.g., per minute) versus long-term price for days or months.
While wireless transmissions and energy consumptions are real-time, energy harvesting depending on ambient environments can operate at an interval of minutes or hours different from the real-time and long-term energy pricing intervals.
To this end, a foresighted plan of energy usage in advance, also known as ``multi-timescale energy planning (MTEP)'', is important to balance energy load and relieve pressure on the grid, and hence improve the reliability and sustainability of the grid.

In this paper, we present a new framework of TWET and MTEP, where stochastic optimization theory is exploited to capture the temporal and spatial randomnesses of both 5G and smart grid in terms of energy price, RES availability, and wireless channel. The minimization of the time-average energy cost of 5G subject to the time-varying harvested RES amounts and energy prices is asymptotically equivalently reformulated to the minimization of a convex closed-form upper bound of the instantaneous cost per timeslot, which in turn can be optimally solved by using convex techniques. This approach is further extrapolated to the case of multiple different timescales, where the instantaneous minimization per timeslot is carried out under the prediction of future RES arrivals and energy prices at longer time horizons based on the past. Extensive simulations show that the energy cost with MTEP can be saved up to 58\%.

In a different yet relevant context, a range of energy-efficient transmission schemes have been developed for 5G systems. In \cite{Wang2013Energy}, a string tautening algorithm was proposed to produce the most energy-efficient schedule for delay-limited traffic, first under the assumption of negligible circuit power, and then extended to non-negligible constant circuit powers \cite{Nan2014Energy,Nan2014Energy1} and energy-harvesting communications \cite{Chen2015Provisioning,Chen2015Energy}. Particularly, coordinated multi-point (CoMP) enables BSs to collaborate for effective inter-cell interference management, thereby substantially improving the energy efficiency of 5G. This technology has been standardized by 3GPP, and extensively studied for energy-efficient applications \cite{Irmer2011Coordinated}. Other active researches on energy-aware wireless techniques are under the way to guaranteeing the quality of service (QoS) of wireless applications using non-persistent RES \cite{Xinchen2017,Qin2017Joint}.

\begin{figure}
\centering
\includegraphics[width=0.45\textwidth]{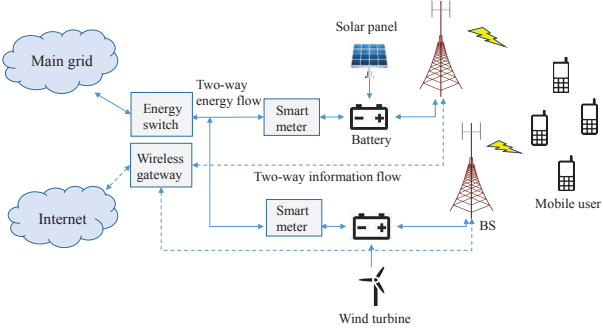}
\vspace{-0.4cm}
\caption{A smart grid powered CoMP system.
Two BSs with local RES harvesting devices and batteries perform TWET with the main grid.
\vspace{-0.4cm}
}
\label{mag_model}
\end{figure}

%\begin{figure}
%\centering
%\includegraphics[width=0.5\textwidth]{Price_new.eps}
%%\vspace{-0.6cm}
%\caption{Multiple timescales of energy pricing and harvesting in the interoperable framework of future 5G and smart grid.}
%%\vspace{-0.6cm}
%\label{multiscale}
%\end{figure}

The rest of this paper is organized as follows. In Section~\ref{model}, the system model is presented. In Sections~\ref{singlescale} and \ref{twoscale}, we propose the new TWET and MTEP algorithms. In Section~V, the asymptotic optimality of the proposed algorithms is established. In Section~VI, numerical results demonstrate the merits of the proposed schemes, followed by conclusions in Section~VII.

\section{System Model}\label{model}

As illustrated in Fig.~\ref{mag_model}, we consider a CoMP downlink system, where a set of ${\cal I}:=\{1,\ldots, I\}$ BSs (e.g., macro/micro/pico BSs) serves a set of ${\cal K}:=\{1,\ldots, K\}$ mobile users.
Each BS is equipped with $M$ transmit antennas, and each user has a single receive antenna. Assume that BSs can harvest RES to support their transmissions. Furthermore, each BS can purchase energy from or sell energy to the grid at time-varying prices via TWET.
Equipped with batteries, the BSs can take advantage of energy price fluctuations, and can store energy for later use.
All BSs are connected to a central controller, which not only collects the communication data from BSs, but also the energy prices via locally installed smart meters.

Suppose that the transmissions are slot-based and experience quasi-static downlink channels, which remain invariant within a slot and vary between slots.
For illustration convenience, the slot duration is normalized to unity, so ``energy'' and ``power'' become interchangeably used throughout this paper.

\subsection{CoMP Transmissions}

Consider a scheduling horizon indexed by the set ${\cal T}:= \{0, \ldots, T-1\}$.
Per slot $t$, let $\mathbf{h}_{ik,t} \in \mathbb{C}^{M}$ denote the vector channel from BS $i$ to user $k$, $\forall i \in {\cal I}$, $\forall k \in \mathcal{K}$; let $\mathbf{h}_{k,t} := [\mathbf{h}_{1k,t}', \ldots, \mathbf{h}_{Ik,t}']'$ collect the channel vectors from all BSs to user $k$, and $\mathbf{H}_t:=[\mathbf{h}_{1,t}. \ldots, \mathbf{h}_{K,t}]$. With linear transmit beamforming performed across the BSs, the vector signal transmitted  to user $k$ is: $\mathbf{q}_k(t) =\mathbf{w}_k(t) s_k(t)$, $\forall k$, where $s_k(t)$ is the information-bearing scalar symbol with unit-energy, and $\mathbf{w}_k(t) \in \mathbb{C}^{MI}$ is the beamforming vector across the BSs serving user $k$. The received vector for user $k$ at slot $t$ is therefore
\begin{align}\label{eq.yk}
y_k(t) =\mathbf{h}_{k,t}^H \mathbf{q}_k(t) + \sum_{l\neq k} \mathbf{h}_{k,t}^H \mathbf{q}_l(t) + n_k(t)
\end{align}
where $\mathbf{h}_{k,t}^H \mathbf{q}_k(t)$ is the desired signal of user $k$, $\sum_{l\neq k} \mathbf{h}_{k,t}^H \mathbf{q}_l(t)$ is the inter-user interference, and $n_k(t)$ is an additive circularly symmetric complex Gaussian (CSCG) noise with zero mean and variance $\sigma_k^2$.

%Let $\mathbf{W}^t:=[\mathbf{w}_1^t,\ldots, \mathbf{w}_K^t]$.
The signal-to-interference-plus-noise ratio (SINR) at user $k$ is given by
\begin{equation}\label{sinr}
    \text{SINR}_k(\{\mathbf{w}_k(t)\})= \frac{|\mathbf{h}_{k,t}^H \mathbf{w}_k(t)|^2}{\sum_{l\neq k} (|\mathbf{h}_{k,t}^H \mathbf{w}_l(t)|^2) + \sigma_k^2}~.
\end{equation}
The transmit power at BS $i$ is given by
\begin{equation}
    P_{x,i}(t) = \sum_{k \in \mathcal{K}} {\mathbf{w}_k^H(t)} \mathbf{B}_i \mathbf{w}_k(t)
\end{equation}
where the matrix $$\mathbf{B}_i:=\text{diag}\left(\underbrace{0, \ldots, 0}_{(i-1)M}, \underbrace{1, \ldots, 1}_{M}, \underbrace{0, \ldots, 0}_{(I-i)M}\right) \in \mathbb{R}^{MI\times MI}$$ selects the corresponding rows out of $\{\mathbf{w}_k(t)\}_{k\in \mathcal{K}}$ to form the $i$-th BS's beamforming vector of size $M\times 1$.

%In practice, the channel state information $\mathbf{h}_k^t$ is seldom accurately available {\it a priori}. Relying on past channel measurements and/or reliable channel predictions, we postulate an additive error model: $\mathbf{h}_k^t = \hat{\mathbf{h}}_k^t + {\bm{\delta}}_k^t$, where $\hat{\mathbf{h}}_k^t$ is the estimated channel known at the BSs. The uncertainty of this estimate is bounded by a spherical region \cite{Vucic2009}:
%\begin{equation}\label{eq.chUncerty}
%  \mathcal{H}_k^t:=\left\{\hat{\mathbf{h}}_k^t + {\bm{\delta}}_k^t \ \biggl| \  \left\|{\bm{\delta}}_k^t \right\| \leq \epsilon_k^t \right\}, \quad \forall k, t
%\end{equation}
%where $\epsilon_k^t>0$ specifies the radius of $\mathcal{H}_k^t$.
%This leads to the worst-case SINR per user $k$ as [cf.~\eqref{sinr}]
%\begin{equation}
%  \widetilde{\text{SINR}}_k (\{\mathbf{w}_k^t\}) := \min_{\mathbf{h}_k^t \in\mathcal{H}_k^t} \frac{|{\mathbf{h}_{k}^t}^H \mathbf{w}_k^t|^2}{\sum_{l\neq k} (|{\mathbf{h}_{k}^t}^H \mathbf{w}_l^t|^2) + \sigma_k^2}~.
%\end{equation}
To guarantee QoS per slot user $k$, the central controller selects a set of $\{\mathbf{w}_k(t)\}$ satisfying [cf. \eqref{sinr}]
\begin{equation}\label{eq.sinr}
    \text{SINR}_k (\{\mathbf{w}_k(t)\}) \geq \gamma_k, \quad \forall k
\end{equation}
where $\gamma_k$ denotes the target SINR value per user $k$.

\subsection{Smart Grid Operations}

Each BS exploits RES.
Let $\mathbf{A}_t:=[A_1(t),\ldots,A_I(t)]'$ collect the harvested RES at slot $t$ across all BSs.
Let $C_i^0$ denote the initial stored energy, and $C_i(t)$ the state of charge of BS $i$ at the beginning of slot $t$.
Each battery has a finite capacity $C_i^{\max}$. A minimum level $C_i^{\min}$ is required at any time for the sake of battery health.
With $P_{b,i}(t)$ denoting the amount of battery charging ($P_{b,i}(t)>0$) or discharging ($P_{b,i}(t)<0$) at slot $t$,
the stored energy then obeys: $\forall i,t,$
\begin{equation}\label{eq.Ci}
    C_i(t+1) = C_i(t)+P_{b,i}(t),\quad C_i^{\min} \leq C_i(t) \leq C_i^{\max}.
\end{equation}
The amount of energy (dis)charged is bounded by
\begin{equation}\label{eq.pb}
     P_{b,i}^{\min} \leq P_{b,i}(t) \leq P_{b,i}^{\max}
\end{equation}
where $P_{b,i}^{\min} <0$ and $P_{b,i}^{\max}>0$.

The total energy consumption $P_{g,i}(t)$ of BS $i$ consists of the beamformers' transmit power $P_{x,i}(t)$, and a constant power $P_c > 0$ consumed by air conditioning, data processor, and circuits. The total consumption $P_{g,i}(t)$ is bounded by $P_{g,i}^{\max}$. Therefore,
\begin{equation}\label{eq.pg}
    P_{g,i}(t)=P_c+P_{x,i}(t)=P_c+\sum_{k \in \mathcal{K}} {\mathbf{w}_k^t}^H \mathbf{B}_i \mathbf{w}_k^t\leq P_{g,i}^{\max}
\end{equation}

The grid can supply $P_{g,i}(t)$ if the harvested RES are insufficient. A BS can also sell its surplus energy (whenever the RES are abundant) back to the grid.
It is clear that the shortage energy of BS $i$ that needs to be purchased from the grid is
$\max\{P_{g,i}(t)-A_i(t)+P_{b,i}(t), 0\}$, while the surplus energy is $\max\{A_i(t)-P_{g,i}(t)-P_{b,i}(t), 0\}$.

With the buying and selling prices $\alpha_t$ and $\beta_t$,
the condition $\alpha_t\geq \beta_t$ holds to avoid meaningless buy-and-sell activities.
%Again, we assume that the prices $(\alpha^t, \beta^t)$ are generated according to an i.i.d. random process for performance analysis purpose,
%while generalization to non-i.i.d. scenarios can be addressed using the techniques in \cite{Geor06}.
Per slot $t$, the energy transaction cost of BS $i$ is therefore given in a convex form by
\begin{align}\label{eq:tranct}
    G(P_{g,i}(t),P_{b,i}(t))& = \max\big\{\alpha_t [P_{g,i}(t)-A_i(t)+P_{b,i}(t)],\notag \\& \beta_t[A_i(t)-P_{g,i}(t)-P_{b,i}(t)]\big\}
\end{align}
Note that at any slot each BS can either purchase electricity from the grid at price $\alpha_t$, or sell surplus to the grid at the price $\beta_t$.
For conciseness, we concatenate all the variables; i.e., $\boldsymbol{\xi}_t:=\{\alpha_t, \beta_t,\mathbf{A}_t, \mathbf{H}_t, \forall t\}$, and suppose that $\boldsymbol{\xi}_t$ is independent and identically distributed (i.i.d.) across time.
%However, $\boldsymbol{\xi}_t$ can be non-i.i.d., and even negatively correlated with each other in practice~\cite{Genoese10}.
%For such a case, it is worth stressing that our proposed algorithm in the sequel can be applied without any modification.
%Yet, performance guarantees in the non-i.i.d. case must be obtained by utilizing the more sophisticated delayed Lyapunov drift techniques along the lines of e.g.~\cite{Geor06}.

\section{Single-Timescale TWET}\label{singlescale}

Over the scheduling horizon ${\cal T}$, the central controller of CoMP seeks the optimal schedule for cooperative transmit beamforming vectors $\{\mathbf{w}_k(t)\}_{k,t}$ and battery (dis)charging energy $\{P_{b,i}(t)\}_{i,t}$, to minimize the total network cost $\sum_{t\in{\cal T}} \sum_{i \in {\cal I}} G(P_{g,i}(t),P_{b,i}(t))$, while guaranteeing QoS, i.e., ${\text{SINR}}_k (\{\mathbf{w}_k(t)\}) \geq \gamma_k$, $\forall k,t$. The problem of interest is formulated as

{\small\begin{align}\label{eq.prob1}
 G^*:=\min_{\{\mathbf{w}_k(t),P_{b,i}(t),C_i^t\}} &\;\lim_{T\rightarrow \infty} \frac{1}{T} \sum_{t=0}^{T-1} \sum_{i \in {\cal I}} \mathbb{E}[G(P_{g,i}(t),P_{b,i}(t))] \\
 \text{s. t.} &~~(\ref{eq.sinr}), (\ref{eq.Ci}), (\ref{eq.pb}), (\ref{eq.pg}), ~~ \forall t,\notag
\end{align}}where the expectation of $G(\cdot)$ is taken over all sources of
randomness.

%It is easy to argue that the objective in (\ref{eq.prob1}) is convex. Indeed, with $\alpha_t > \beta_t$, the transaction cost can be alternatively written as
%\begin{align}\label{eq.Gr}
%    G(P_{g,i}(t),P_{b,i}(t))& = \max\big\{\alpha_t [P_{g,i}(t)-A_i(t)+P_{b,i}(t)],\notag \\& \beta_t[A_i(t)-P_{g,i}(t)-P_{b,i}(t)]\big\}
%\end{align}
%which is clearly convex \cite{convex}; and so is the objective in (\ref{eq.prob1}).

The SINR constraints in (\ref{eq.sinr}) can be rewritten into convex second-order cone (SOC) constraints through proper rearrangement \cite{Wiesel2005Linear}; that is,
\begin{equation}\label{sinr-trans}
\begin{split}
    & \sqrt{\sum_{l \neq k} |\mathbf{h}_{k,t}^H \mathbf{w}_l(t)|^2 + \sigma_k^2} \leq \frac{1}{\sqrt{\gamma_k}} \text{Re}\{\mathbf{h}_{k,t}^H \mathbf{w}_k(t)\}, \\
    &\text{Im}\{\mathbf{h}_{k,t}^H \mathbf{w}_k(t)\} =0, ~~\forall k.
\end{split}
\end{equation}

Although (\ref{eq.prob1}) is convex based on \eqref{eq:tranct} and \eqref{sinr-trans},
it is difficult to solve since the minimization of
average total cost is over the infinite time horizon. Particularly, the (dis)charging operations are coupled across time through the battery level changes. The decision at current slot can affect the decisions further down into the future.

By recognizing that (\ref{eq.Ci}) can be interpreted as an energy queue recursion, we apply the time decoupling technique to turn \eqref{eq.prob1} into a tractable form~\cite{Urgaonkar2011Optimal}. Over the infinite time horizon, \eqref{eq.Ci} can be replaced by
\begin{equation}\label{eq.Pb0}
\lim_{T\rightarrow \infty} \frac{1}{T} \sum_{t=0}^{T-1} \mathbb{E}[P_{b,i}(t)]=0,~~\forall i.
\end{equation}

We can then resort to the Lyapunov optimization method \cite{neely2010stochastic} to achieve asymptotically optimal solution for (\ref{eq.prob1}). Introducing two critical parameters, namely a ``queue perturbation'' parameter $\Gamma$ and a weight parameter $V$, the problem to be solved becomes
\begin{align}\label{eq.rtpro}
        \min_{\{\mathbf{w}_k(t),P_{b,i}(t)\}} \;& \sum_{i\in \cal I} \biggl\{V G(P_{g,i}(t),P_{b,i}(t))\biggr.
         \biggl.+ Q_i(t) P_{b,i}(t)\biggr\}\nonumber\\
        \text{s. t.}\; & ~~(\ref{eq.pb}), (\ref{eq.pg}), (\ref{sinr-trans}),
        %&~P_c+\sum_k {\mathbf{w}_k^H(t)} \mathbf{B}_i \mathbf{w}_k(t) + P_{b,i}(t) \leq \frac{E_i^*[n_t]}{T} + P_i(t), ~\forall i
\end{align}
where we define a virtual queue $Q_i(t) := C_i(t) +\Gamma$, $\forall i, t$.

Given that $G(\cdot)$ is convex and increasing, $G(P_{g,i}(t),P_{b,i}(t))$ is jointly convex in $(\mathbf{w}_k(t), P_{b,i}(t))$ \cite[Sec.~3.2.4]{convex}. It readily follows that (\ref{eq.rtpro}) is a convex optimization problem, and can be solved via off-the-shelf solvers, as summarized in Algorithm~1.

%%%%%%%%%%%%%%%%%%%%%%%%%%%%%%%%%%%%%%%%%%%%%%%%%%%%%
\begin{algorithm}[t]
\caption{Two-Way Energy Trading (TWET)}\label{algo:TWET}
\begin{algorithmic}[1]
\State \textbf{Initialization}: Select $\Gamma$ and $V$, and introduce a virtual queue $Q_i(0) := C_i(0) +\Gamma$, $\forall i$.

\State \textbf{Energy trading and beamforming}: At every slot $t$, observe a realization $\boldsymbol{\xi}_t$, and decide $\{\mathbf{w}_k^*(t), \forall k; P_{b,i}^*(t), \forall i\}$ by solving \eqref{eq.rtpro}.
The BSs perform two-way energy trading with the main grid and coordinated beamforming based on $\{\mathbf{w}_k^*(t), \forall k; P_{b,i}^*(t), \forall i\}$.
\State \textbf{Queue updates}: Per slot $t$, (dis)charge the battery based on $\{P_{b,i}^*(t), \forall i\}$, so that the stored energy $C_i(t+1) = C_i(t) + P_{b,i}^*(t)$, $\forall i$; and update the virtual queues $Q_i(t+1) := C_i(t+1) +\Gamma, \forall i$.
\end{algorithmic}
\end{algorithm}
%%%%%%%%%%%%%%%%%%%%%%%%%%%%%%%%%%%%%%%%%%%%%%%%%%%%%%

\section{Multi-Timescale MTEP}\label{twoscale}

A more general case of energy trading between 5G and
smart grid can involve multiple timescales of
real-time wireless transmission, energy harvesting, and short-term/long-term energy pricing, as discussed earlier. Long-term energy buying prices are lower than real-time prices on average. This discrepancy can be
further exploited to reduce the energy cost of 5G, as compared
to TWET. To this end, MTEP is expected to plan energy usage and purchase ahead-of-time over multiple timescales. For illustration
convenience, here we consider two different timescales, i.e., real-time for a short slot and ahead-of-time for a long interval. The wireless transmission is synchronized with the real-time energy pricing, while the output of energy harvesting is synchronized with the ahead-of-time energy pricing.

At the beginning of each ``coarse-grained'' interval, namely at time $t=nT$, $n=1, 2, \ldots$, $A_{i,n}$ denotes the RES amount collected per BS $i \in {\cal I}$, and $\mathbf{A}_n:=[A_1[n],\ldots,A_I[n]]'$. Given $\mathbf{A}_n$, the central controller decides the energy amount $E_i[n]$ to be used in the next $T$ slots per BS $i$, i.e., the grid supplies an average energy amount of $E_i[n]/T$ per slot $t=nT, \ldots, (n+1)T-1$.
Given $E_i[n]$ and $A_i[n]$, the shortage energy purchased from the grid for BS $i$ is $\max\{E_i[n]-A_i[n], 0\}$; or the surplus energy sold to the grid is $\max\{A_i[n]-E_i[n], 0\}$.
BS $i$ either buys energy from the grid at the ahead-of-time price $\alpha_n^{{\rm lt}}$, or sells energy to the grid at price $\beta_n^{{\rm lt}}$.
We set $\alpha_n^{{\rm lt}} > \beta_n^{{\rm lt}}$ to avoid meaningless buy-and-sell activities. The ahead-of-time energy transaction cost of BS~$i$ is therefore given by
%Suppose that the energy can be purchased from the grid at the ahead-of-time (i.e., long-term) price $\alpha_n^{{\rm lt}}$, while the energy can be sold to the grid at price $\beta_n^{{\rm lt}}$. Notwithstanding, we shall always set $\alpha_n^{{\rm lt}} > \beta_n^{{\rm lt}}$  to avoid meaningless buy-and-sell activities of the BSs for profit. The transaction cost with BS $i$ for such an energy planning is therefore given by
\begin{equation}
\label{eq.lt-tranct}
G^{{\rm lt}}(E_i[n]):= \max\{\alpha_n^{{\rm lt}}(E_i[n]\!-\!A_i[n]), \beta_n^{{\rm lt}} (A_i[n]\!-\!E_i[n])\}.
\end{equation}
Let $\boldsymbol{\xi}_n^{\rm lt}:=\{\alpha_n^{\rm lt}, \beta_n^{\rm lt},\mathbf{A}_n, \forall n\}$ collect all the random variables evolving at this slow timescale.

With $n_t := \lfloor \frac{t}{T} \rfloor$, the real-time energy buying and selling prices $\alpha_t^{\rm rt}$ and $\beta_t^{\rm rt}$, and $A_i(t)$ in \eqref{eq:tranct} substituted by $\frac{E_i[n_t]}{T}$, the real-time energy transaction cost of BS $i$ is given by

%
%Let $P_i(t)$ denote the real-time energy amount that is purchased from ($P_i(t) >0$) or sold to ($P_i(t) <0$) the grid by BS $i$. Let $\alpha_t^{\rm rt}$ and $\beta_t^{{\rm rt}}$ ($\alpha_t^{\rm rt} > \beta_t^{{\rm rt}}$) denote the real-time energy purchase and selling prices, respectively. Then the real-time energy transaction cost for BS $i$ is
\begin{align}\label{eq:rt-tranct}
&G^{{\rm rt}}(E_i[n_t],P_{g,i}(t),P_{b,i}(t)):=
\max\big\{\alpha_t^{\rm rt} [P_{g,i}(t)\notag\\
&-\frac{E_i[n_t]}{T}+P_{b,i}(t)], \beta_t^{\rm rt}[\frac{E_i[n_t]}{T}-P_{g,i}(t)-P_{b,i}(t)]\big\}
\end{align}
Let $\boldsymbol{\xi}_t^{\rm rt}:=\{\alpha_t^{\rm rt}, \beta_t^{\rm rt},\mathbf{H}_t, \forall t\}$ collect all the random variables evolving at this fast timescale.
%\begin{equation}
%\label{eq.rt-tranct}
%G^{{\rm rt}}(P_i(t)):= \alpha_t^{\rm rt}[P_i(t)]^{+} - \beta_t^{{\rm rt}} [-P_i(t)]^{+}.
%\end{equation}
%With $n_t := \lfloor \frac{t}{T} \rfloor$, we have the following demand-and-supply balance equation per slot $t$:
%\begin{equation}\label{eq.balance}
%   P_c+\sum_{k \in \mathcal{K}} {\mathbf{w}_k^H(t)} \mathbf{B}_i \mathbf{w}_k(t) + P_{b,i}(t) = \frac{E_i[n_t]}{T} + P_i(t), \;\forall i.
%\end{equation}

According to (\ref{eq.lt-tranct}) and (\ref{eq:rt-tranct}), we can define the energy transaction cost for BS $i$ per slot $t$ as:
\begin{equation}\label{eq.gt}
    \Phi_i(t):=\frac{1}{T} G^{{\rm lt}}(E_i[n_t]) + G^{{\rm rt}}(E_i[n_t],P_{g,i}(t),P_{b,i}(t)).
\end{equation}
Let ${\cal X}:=\{E_i[n], \forall i,n; P_{b,i}(t), C_i(t), \forall i, t; \mathbf{w}_k(t), \forall k, t\}$. The goal is to design an online energy management scheme that decides the ahead-of-time energy-trading amounts $\{E_i[n], \forall i\}$ at every $t=nT$, battery (dis)charging amounts $\{P_{b,i}(t), \forall i\}$, and the CoMP beamforming vectors $\{\mathbf{w}_k(t),\forall k\}$ per slot $t$, so as to minimize the total time-average energy cost. The problem of MTEP is to find
\begin{equation}\label{eq.prob}
\begin{split}
   \Phi^* := & \min_{{\cal X}} \;\lim_{N\rightarrow \infty} \frac{1}{NT} \sum_{t=0}^{NT-1} \sum_{i \in {\cal I}} \mathbb{E} \{\Phi_i(t)\} \\
   & \text{subject to} ~~~ (\ref{eq.Ci}), (\ref{eq.pb}), (\ref{eq.pg}), (\ref{sinr-trans}),~~ \forall t
\end{split}
\end{equation}
where $\Phi^*$ is the optimal solution, and the expectations of $\Phi_i(t)$ are taken over all sources of randomness.
%Note that here the constraints (\ref{eq.sinr}),  (\ref{eq.Ci}),  (\ref{eq.pb}) and (\ref{eq.pg}) are implicitly required to hold for every realization of the underlying random states $\boldsymbol{\xi}_t^{\rm rt}$ and $\boldsymbol{\xi}_n^{\rm lt}$.

We can generalize the Lyapunov optimization techniques to achieve the asymptotically optimal solution for (\ref{eq.prob}). Both ahead-of-time and real-time decisions on energy trading can be accommodated. This consists of multiple asynchronous Lyapunov optimization processes running at different timescales.

Problem (\ref{eq.prob}) can be decoupled into two subproblems.
One is real-time energy trading and beamforming at each fine-grained time slot within a coarse-grained interval $n$, given the energy plan made at the beginning of the interval $E_i[n]$. Since \eqref{eq.prob} recedes to \eqref{eq.prob1}, we can just run Algorithm 1 with $A_i(t)=\frac{E_i[n_t]}{T}$ in \eqref{eq.rtpro}. The other subproblem is ahead-of-time energy planning of the optimal $E_i^*[n]$, as given by
\begin{align}\label{eq.lt-prob}
       \min_{\{E_i[n]\}} \; &\sum_{i\in \cal I} \Big\{V\big[G^{{\rm lt}}(E_i[n])+\sum_{t = \tau}^{\tau+T-1} \mathbb{E}\{G^{{\rm rt}}(E_i[n],\nonumber\\
       & \quad P_{g,i}(t),P_{b,i}(t))\}\big]
        +\sum_{t = \tau}^{\tau+T-1} Q_i(\tau)\mathbb{E}\{P_{b,i}(t)\}\Big\}\nonumber\\
      & \text{s. t.} ~(\ref{eq.pb}), (\ref{eq.pg}), (\ref{sinr-trans}), ~\forall t =\tau, \ldots, \tau+T-1\!
\end{align}
where the expectations are taken over $\boldsymbol{\xi}_t^{\rm rt}$.

%\subsection{Real-Time Energy Balancing and Beamforming}
%
%It is easy to argue that the objective (\ref{eq.rt-prob}) is convex. Indeed, with $\alpha_t^{\rm rt} > \beta_t^{{\rm rt}}$, the transaction cost with $P_i(t)$ can be alternatively written as
%\begin{equation}\label{eq.Gr}
%    G^{{\rm rt}}(P_i(t)) = \max\{\alpha_t^{\rm rt} P_i(t), \; \beta_t^{{\rm rt}} P_i(t)\}
%\end{equation}
%which is clearly convex \cite{convex}; and so is the objective in (\ref{eq.rt-prob}).
%
%
%We can rewrite the problem (\ref{eq.rt-prob}) as
%\begin{align}\label{eq.rt-prob1}
%        \min \; & \sum_{i\in \cal I} \biggl\{V G^{{\rm rt}}(P_c+\sum_{k\in \cal K} {\mathbf{w}_k^H(t)} \mathbf{B}_i \mathbf{w}_k(t) + P_{b,i}(t) - \frac{E_i^*[n_t]}{T})\biggr.\nonumber\\
%         & ~~~~~~~~~~ \biggl.+ Q_i(n_tT) P_{b,i}(t)\}\biggr\}\nonumber\\
%        \text{s. t.}\; & (\ref{eq.pb}), (\ref{eq.pg}), (\ref{sinr-trans}).
%\end{align}
%As $G^{{\rm rt}}(\cdot)$ is convex and increasing, it is easy to see that $G^{{\rm rt}}(P_c+\sum_k {\mathbf{w}_k^H(t)} \mathbf{B}_i \mathbf{w}_k(t) + P_{b,i}(t) - E_i^*[n_t]/T)$ is jointly convex in $(P_{b,i}(t), \{\mathbf{w}_k(t)\})$ \cite[Sec.~3.2.4]{convex}. It then readily follows that (\ref{eq.rt-prob1}) is a convex optimization problem, which can be solved via off-the-shelf solvers.

%To solve (\ref{eq.lt-prob}), the probability distribution function (pdf) of the random state $\boldsymbol{\xi}_t^{\rm rt}$ must be known across slots $t = nT, \ldots, (n+1)T-1$. However, this pdf is seldom available in practice.
Next we develop an efficient solver of (\ref{eq.lt-prob}).
Suppose that $\boldsymbol{\xi}_t^{\rm rt}$ is i.i.d. across time slots. We can suppress $t$ from all optimization variables, and rewrite (\ref{eq.lt-prob}) as (with short-hand notation $Q_i[n]:=Q_i(nT)$)

\begin{subequations}
\label{eq.lt-prob1}
{\small \begin{align}
\min_{\{E_i{[n]}\}} & \sum_{i \in \cal I} \{V G^{{\rm lt}}(E_i[n])+T \mathbb{E}[VG^{{\rm rt}}(E_i[n],P_{g,i}(\boldsymbol{\xi}_t^{\rm rt}),P_{b,i}(\boldsymbol{\xi}_t^{\rm rt}))\nonumber\\
&+ Q_i[n]P_{b,i}(\boldsymbol{\xi}_t^{\rm rt})]\}\nonumber \\
\text{s. t.} ~
& \sqrt{\sum_{l \neq k} |\mathbf{h}_{k}^H \mathbf{w}_l(\boldsymbol{\xi}_t^{\rm rt})|^2 + \sigma_k^2} \leq \frac{1}{\sqrt{\gamma_k}} \text{Re}\{\mathbf{h}_{k}^H \mathbf{w}_k(\boldsymbol{\xi}_t^{\rm rt})\}, \nonumber\\
& \text{Im}\{\mathbf{h}_{k}^H \mathbf{w}_k(\boldsymbol{\xi}_t^{\rm rt})\} =0,~~\forall k, \boldsymbol{\xi}_t^{\rm rt}  \label{eq.p1a}\\
& P_{b}^{\min} \leq P_{b,i}(\boldsymbol{\xi}_t^{\rm rt}) \leq P_{b}^{\max},~~ \forall i, \boldsymbol{\xi}_t^{\rm rt} \label{eq.p1b}\\
& P_c+\sum_{k\in \cal K} {\mathbf{w}_k^H(\boldsymbol{\xi}_t^{\rm rt})} \mathbf{B}_i \mathbf{w}_k(\boldsymbol{\xi}_t^{\rm rt}) \leq P_g^{\max}, ~~\forall i, \boldsymbol{\xi}_t^{\rm rt},\label{eq.p1c}
\end{align}}\end{subequations}which can be further reformulated as an unconstrained optimization problem with respect to $E_i[n]$, as given by
%Note that this form explicitly indicates the dependence of the decision variables $\{\mathbf{w}_k, P_{b,i}\}$ on the realization of $\boldsymbol{\xi}_t^{\rm rt}$.
%Since the energy planning problem (\ref{eq.lt-prob}) only determines the optimal ahead-of-time energy purchase $E_i^*[n]$, we can rewrite \eqref{eq.lt-prob1}

\begin{equation}\label{eq.lt-prob2}
     \min_{\{E_i[n]\}} \; \sum_{i \in \cal I} \Big[V G^{{\rm lt}}(E_i[n]) + T\bar{G}^{{\rm rt}}(\{E_i[n]\})\Big],
\end{equation}
where we define
\begin{align}\label{eq.Grt}
   \!\!\!\bar{G}^{{\rm rt}}\!(\!\{E_i[n]\}\!)\!&\!:=\!\!\!\! \min_{\{\!P_i,P_{b,i},\mathbf{w}_k\!\}} \!\sum_{i \in \cal I}\mathbb{E}\biggl\{\!V\Psi^{\rm rt}\!(E_i[n],P_{b,i}(\boldsymbol{\xi}_t^{\rm rt}),\mathbf{w}_k(\boldsymbol{\xi}_t^{\rm rt}))\nonumber\\
    &\!+\!Q_i[n]P_{b,i}(\boldsymbol{\xi}_t^{\rm rt})\biggl\} ~\text{s. t. (\ref{eq.p1a}), (\ref{eq.p1b}), (\ref{eq.p1c})}
\end{align}
with the compact notation $\Psi^{{\rm rt}}(E_i[n], \mathbf{w}_k(\boldsymbol{\xi}_t^{\rm rt}), P_{b,i}(\boldsymbol{\xi}_t^{\rm rt})):=  G^{{\rm rt}}(E_i[n],P_{g,i}(\boldsymbol{\xi}_t^{\rm rt}),P_{b,i}(\boldsymbol{\xi}_t^{\rm rt}))$.

Since $\mathbb{E}[V\Psi^{{\rm rt}}(E_i[n], \mathbf{w}_k(\boldsymbol{\xi}_t^{\rm rt}), P_{b,i}(\boldsymbol{\xi}_t^{\rm rt}))+ Q_i[n]P_{b,i}(\boldsymbol{\xi}_t^{\rm rt})]$ is jointly convex in $(E_i, \mathbf{w}_k, P_{b,i})$, then the minimization over $(\mathbf{w}_k, P_{b,i})$ is within a convex set; thus, \eqref{eq.p1a}-(\ref{eq.p1c}) are convex with respect to $E_i[n]$ \cite[Sec.~3.2.5]{convex}.
Hence, \eqref{eq.lt-prob2} is generally a nonsmooth and unconstrained convex problem, and can be solved using the stochastic subgradient method.

The subgradient of $G^{{\rm lt}}(E_i[n])$ can be written as
\[
    \partial G^{{\rm lt}}(E_i[n]) =
    \begin{cases}
        \alpha_n^{{\rm lt}}, ~~\text{if} ~ E_i[n]> A_i[n]; \\
        \beta_n^{{\rm lt}},  ~~\text{if} ~ E_i[n]< A_i[n]; \\
        \text{any } x \in [\beta_n^{{\rm lt}}, \alpha_n^{{\rm lt}}], ~~\text{if} ~ E_i[n]=A_i[n].
    \end{cases}
\]
With the optimal solution $\{\mathbf{w}_k^{E}(\boldsymbol{\xi}_t^{\rm rt}),P_{b,i}^{E}(\boldsymbol{\xi}_t^{\rm rt})\}$ for (\ref{eq.Grt}), the partial subgradient of $\bar{G}^{{\rm rt}}(\{E_i[n]\})$ with respect to $E_i[n]$ is $\partial_i \bar{G}^{{\rm rt}}(\{E_i[n]\}) = V \mathbb{E}\{\partial \Psi^{{\rm rt}}(E_i[n], \mathbf{w}_k^{E}(\boldsymbol{\xi}_t^{\rm rt}), P_{b,i}^{E}(\boldsymbol{\xi}_t^{\rm rt}))\}$, where
\[
    \partial \Psi^{{\rm rt}}(E_i[n], \mathbf{w}_k^{E}(\boldsymbol{\xi}_t^{\rm rt}), P_{b,i}^{E}(\boldsymbol{\xi}_t^{\rm rt})) =
    \begin{cases}
        \frac{-\beta_t^{{\rm rt}}}{T}, ~ \text{if}  ~\frac{E_i[n]}{T}> \Delta; \\
        \frac{-\alpha_t^{\rm rt}}{T}, ~ \text{if} ~\frac{E_i[n]}{T} <\Delta; \\
        x \in [\frac{-\alpha_t^{\rm rt}}{T}, \frac{-\beta_t^{{\rm rt}}}{T}], ~\text{else},
    \end{cases}
\]
and $\Delta:=P_c+\sum_k {{\mathbf{w}_k^{E}}^H(\boldsymbol{\xi}_t^{\rm rt})} \mathbf{B}_i \mathbf{w}_k^{E}(\boldsymbol{\xi}_t^{\rm rt})+ P_{b,i}^{E}(\boldsymbol{\xi}_t^{\rm rt})$.

Let $\bar{g_i}(E_i):=V\partial G^{{\rm lt}}(E_i)+T\partial_i \bar{G}^{{\rm rt}}(\{E_i\})$. The sub-gradient descent iteration can be employed to find the optimal $E_i^*[n]$ for (\ref{eq.lt-prob2}), as
\begin{equation}\label{eq.sub}
    E_i^{(j+1)}[n] = [E_i^{(j)}[n] - \mu^{(j)} \bar{g}_i(E_i^{(j)}[n])]^+, \quad \forall i
\end{equation}
where $j$ denotes iteration index, and $\{\mu^{(j)}\}$ collects stepsizes.

Since the distribution of $\boldsymbol{\xi}_t^{\rm rt}$ is unknown a priori,
a stochastic subgradient approach is derived based on the past realizations $\{\boldsymbol{\xi}_{\tau}^{\rm rt}, \tau =0,1,\ldots, nT-1\}$.
We can randomly draw a realization $\boldsymbol{\xi}_{\tau}^{\rm rt}$ from past realizations, and run the following iteration
\begin{equation}\label{eq.stocsub}
    E_i^{(j+1)}[n] = [E_i^{(j)}[n] - \mu^{(j)} g_i(E_i^{(j)}[n])]^+, \quad \forall i
\end{equation}
where $g_i(E_i^{(j)}[n]):=V(\partial G^{{\rm lt}}(E_i^{(j)}[n])+ T\partial \Psi^{{\rm rt}}(E_i^{(j)}[n], \linebreak \mathbf{w}_k^{E}(\boldsymbol{\xi}_{\tau}^{\rm rt}), P_{b,i}^{E}(\boldsymbol{\xi}_{\tau}^{\rm rt}) ))$ with
$\{\mathbf{w}_k^{E}(\boldsymbol{\xi}_{\tau}^{\rm rt}), P_{b,i}^{E}(\boldsymbol{\xi}_{\tau}^{\rm rt})\}$ obtained by solving a convex problem \eqref{eq.Grt} with $E_i[n] =E_i^{(j)}[n]$. The iteration (\ref{eq.stocsub}) can asymptotically converge to the optimal $\{E_i^*[n], \forall i\}$ as $j \rightarrow \infty$ \cite{convex}.

The proposed MTEP is presented in Algorithm~2.

\begin{algorithm}[t]
\caption{Multi-Timescale Energy Planning (MTEP)}\label{algo:MTEP}
\begin{algorithmic}[1]
\State \textbf{Initialization}: Select $\Gamma$ and $V$, and introduce a virtual queue $Q_i(0) := C_i(0) +\Gamma$, $\forall i$.

\State \textbf{Ahead-of-time energy planning}: Per interval $\tau=nT$, observe a realization $\boldsymbol{\xi}_n^{\rm lt}$, and determine the energy amounts $\{E_i^*[n], \forall i\}$ by solving \eqref{eq.lt-prob}.
Then the BSs trade energy with the main grid based on $\{E_i^*[n], \forall i\}$, and request the grid to supply an average amount $E_i^*[n]/T$ per slot $t = \tau, \ldots, \tau+T-1$.
\State \textbf{Energy trading and beamforming}: Run Algorithm 1 with $A_i(t)=\frac{E_i[n_t]}{T}$ in \eqref{eq.rtpro}.
\State \textbf{Queue updates}: Per slot $t$, (dis)charge the battery based on $\{P_{b,i}^*(t), \forall i\}$, so that the stored energy $C_i(t+1) = C_i(t) + P_{b,i}^*(t)$, $\forall i$; and update the virtual queues $Q_i(t+1) := C_i(t+1) +\Gamma, \forall i$.
\end{algorithmic}
\end{algorithm}

\section{Asymptotic Optimality of TWET and MTEP}

The asymptotic optimality of the proposed Algorithms 1 and 2 can be established through the following theorems.

\textbf{Theorem 1 \cite{wang2016dynamic}}.
If we set $Q_i^0 = C_i^0 - V\bar{\alpha} + P_{b,i}^{\min} - C_i^{\min}$, $\forall i$, and select a $V \leq V^{\max}$,
then the TWET algorithm yields a feasible dynamic control scheme for~\eqref{eq.prob1}, which is asymptotically optimal in the sense that
\[
     G^* \leq \lim_{T\rightarrow \infty} \frac{1}{T} \sum_{t=0}^{T-1} \sum_{i \in {\cal I}} \mathbb{E}[\hat{G}(t)] \leq G^* + \frac{M_1}{V}
\]
where $\hat{G}(t)$ denotes the resultant cost under the proposed TWET algorithm; $\bar{\alpha}:=\max\{\alpha_t, \forall t\}$, $\underline{\beta}:=\min\{\beta_t, \forall t\}$, $M_1 := \frac{1}{2}\sum_i (\max\{P_{b,i}^{\max}, -P_{b,i}^{\min}\})^2$, and $V^{\max}:= \min_i \{C_i^{\max}  - C_i^{\min}+ P_{b,i}^{\min} - P_{b,i}^{\max}\}/(\bar{\alpha}-\underline{\beta})$.

\textbf{Theorem 2 \cite{wang2016two}}.
If we set $Q_i^0 = C_i^0 - V\bar{\alpha} + TP_{b,i}^{\min} - C_i^{\min}$, $\forall i$, and select a $V \leq V^{\max}$, Then the proposed MTEP yields a feasible dynamic control scheme for~\eqref{eq.prob}, which is asymptotically near-optimal in the sense that
\[
    \Phi^* \leq \lim_{N\rightarrow \infty} \frac{1}{NT} \sum_{t=0}^{NT-1} \sum_{i \in {\cal I}} \mathbb{E} [\hat{\Phi_i}(t)] \leq \Phi^* + \frac{M_2}{V}
\]
where $\hat{\Phi_i}(t)$ denotes the resultant cost under the proposed MTEP algorithm; $\bar{\alpha}:=\max\{\alpha_t^{\rm rt}, \forall t\}$, $\underline{\beta}:=\min\{\beta_t^{\rm rt}, \forall t\}$, $M_2 := \frac{T}{2}\sum_i (\max\{P_{b,i}^{\max}, -P_{b,i}^{\min}\})^2$, and $V^{\max}:= \min_i \{C_i^{\max}  - C_i^{\min}+ T(P_{b,i}^{\min} - P_{b,i}^{\max})\}/(\bar{\alpha}-\underline{\beta})$.

Clearly, Algorithms 1 and 2 provide solutions as close to the optimum $G^*$ or $\Phi^*$ as possible, if we have very small price difference $(\bar{\alpha}-\underline{\beta})$ or very large battery capacities $\{C_i^{\max}\}_i$.

\section{Numerical Tests}\label{simulation}

Comparison studies are conducted between Algorithms 1 and 2, as well as an offline
scheme \cite{Wang2015Robust} to benchmark their performances. Note that the offline benchmark is an ideal scheme with a-priori knowledge of future channel states, energy prices and RES arrivals. We also simulate a heuristic algorithm (Heu) that minimizes the instantaneous energy cost per slot without battery (dis)charging.

We considered two BSs, each with two transmit antennas, to serve three single-antenna mobile users. The system bandwidth is 1~MHz. For simplicity, we randomly generate the channel coefficients as zero-mean complex-Gaussian random variables with unit variance. We set $P_{b,i}^{\max}=2$ kWh, $P_{b,i}^{\min}=-2$ kWh, $C_i^{\max}=60$ kWh, and $C_i^{\min}=0$. The SINR requirement is $\gamma_k^{\text{req}}= 5$ dB for all users (unless otherwise specified). A coarse-grained interval consists of $T=5$ time slots.
The ahead-of-time and real-time energy buying prices $\alpha_n^{{\rm lt}}$ and $\alpha_t^{\rm rt}$ follow folded normal distributions with $\mathbb{E}\{\alpha_n^{{\rm lt}}\}=~$\$1.5/kWh and $\mathbb{E}\{\alpha_t^{\rm rt}\}=~ $\$2.3/kWh. The corresponding selling prices are $\beta_n^{{\rm lt}}=0.9\times \alpha_n^{{\rm lt}}$ and $\beta_t^{{\rm rt}}=0.3 \times\alpha_t^{\rm rt}$.
The harvested energy $A_i[n]$ also yield from a folded normal distribution, with an average rate of 1.6 kWh/slot.

\begin{figure}[t]
\centering
\includegraphics[width=0.36\textwidth]{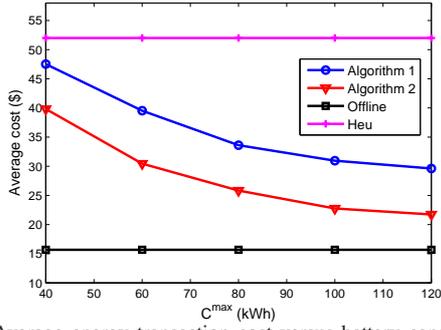}
\vspace{-0.4cm}
\caption{Average energy transaction cost versus battery capacity $C_i^{\max}$.}
\label{fig: Cmax}
\vspace{-0.4cm}
\end{figure}

The average transaction costs of the four algorithms are depicted under different battery capacities $C_i^{\max}$ in Fig.~\ref{fig: Cmax}. Clearly, the growth of $C_i^{\max}$ from 40 to 120 has no impact on the average costs of the offline scheme and the Heu, but causes the other two to monotonically decrease. In particular, reductions of 42\% and 58\% can be achieved using Algorithms~1 and 2 when $C_i^{\max}=120$ kWh, respectively, as compared to the Heu. Algorithm 2 always outperforms Algorithm 1 and the Heu, since Algorithm 2 is able to take advantage of multi-timescale energy pricing, while Algorithm 1 can only work with real-time prices. The Heu without battery (dis)charging has to purchase much more expensive energy from the real-time energy market, thus resulting in the highest transaction cost.

\begin{figure}[t]
\centering
\includegraphics[width=0.36\textwidth]{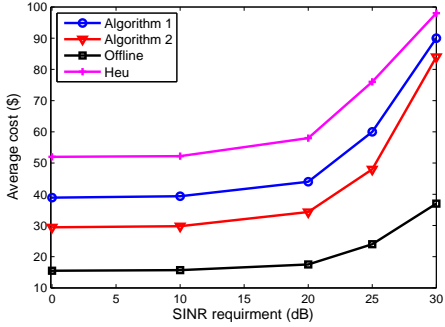}
\vspace{-0.3cm}
\caption{Average energy transaction cost versus SINR requirements.}
\label{fig: QoS}
\vspace{-0.4cm}
\end{figure}

Fig.~\ref{fig: QoS} compares the average energy costs of the four algorithms under different SINR requirements $\gamma_k^{\text{req}}$. It can be seen that all the average costs increase with the growth of $\gamma_k^{\text{req}}$ (which represents more strict SINR requirements). Again, Algorithm 2 outperforms Algorithm 1 and the Heu, incurring a cost closest to the offline benchmark, given the same SINR requirement.
However, note that the optimal offline counterpart cannot work in practice due to the lack of future stochastic system information.

\begin{figure}[t]
\centering
\includegraphics[width=0.36\textwidth]{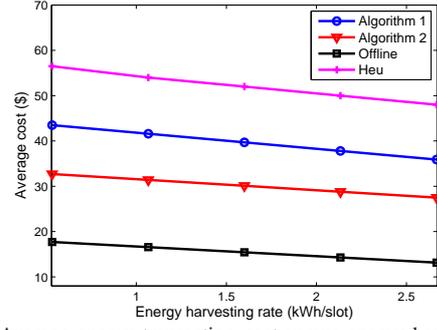}
\vspace{-0.4cm}
\caption{Average energy transaction cost versus energy harvesting rate. }
\label{fig: BS}
\vspace{-0.4cm}
\end{figure}

Fig.~\ref{fig: BS} plots the average energy costs against the energy
harvesting rate (in kWh/slot). We can see that the costs decline
linearly with the growth of energy harvesting capability. Also, it can be observed that the gap between
Algorithm 2 and the offline benchmark remains almost
unchanged; while that between Algorithms 1 and 2 decreases
as the energy harvesting rate increases. We can conclude that the long-term energy planning in MTEP is particularly effective to
the systems with limited energy harvesting capability.

\section{Conclusion}

In this article, we present a new framework of TWET and
MTEP, where Lyapunov optimization techniques are exploited to
capture the temporal and spatial randomnesses of both 5G
and smart grid in terms of energy price, RES, and wireless channel. Simulation results show that effective MTEP is able to save 58\% of the energy cost of 5G.

\end{document}